\documentclass[aip,
sd,%
amsmath,amssymb,
preprint,
]{revtex4-1}
\usepackage{graphicx}
\usepackage{amsmath}
\usepackage[update,prepend]{epstopdf}
\usepackage{dcolumn}
\usepackage{bm}

\begin{document}

\title{Identification of spin wave modes strongly coupled to a co-axial cavity.}

\author{N. J. Lambert}
\affiliation{Microelectronics Group, Cavendish Laboratory, University of Cambridge, Cambridge, CB3 0HE, UK}
\author{J. A. Haigh}
\affiliation{Hitachi Cambridge Laboratory, Cavendish Laboratory, University of Cambridge, Cambridge, CB3 0HE, UK}
\author{A. J. Ferguson}
\affiliation{Microelectronics Group, Cavendish Laboratory, University of Cambridge, Cambridge, CB3 0HE, UK}
\email{ajf1006@cam.ac.uk}

\date{\today}

\begin{abstract}
We demonstrate, at room temperature, the strong coupling of the fundamental and non-uniform magnetostatic modes of an yttrium iron garnet (YIG) ferrimagnetic sphere to the electromagnetic modes of a co-axial cavity. The well-defined field profile within the cavity yields a specific coupling strength for each magnetostatic mode. We experimentally measure the coupling strength for the different magnetostatic modes and, by calculating the expected coupling strengths, are able to identify the modes themselves.

\end{abstract}

\maketitle

A magnet may be excited in a uniform mode\cite{Kittel1947,Kittel1948}, where all the constituent moments are precessing in phase, or in non-uniform modes\cite{Walker1957,Fletcher1959} where there is a spatially varying phase difference between the moments. The uniform oscillating field that usually drives ferromagnetic resonance excites only the uniform mode or higher order modes with a net dynamic magnetisation. In contrast, if the oscillating field is spatially dependent, perhaps due to the skin depth in the case of a metal ferromagnet\cite{Kittel1958,Seavey1958} or by design in an electromagnetic waveguide or cavity\cite{Khivintsev2010,Goryachev2014a}, then the modes are excited according to the spatial symmetry of the drive field. Such modes are the standing spin waves, and their propagating counterparts are central to the research field of magnonics which introduces the possibility to transfer information over millimeter length scales\cite{Eshbach1962,Tsoi2000} and perform specific information processing tasks\cite{Chumak2014}. 

Recently there has been a surge of interest in the coupling of magnets to high quality factor electromagnetic cavities\cite{Soykal2010,Soykal2010a}, motivated by the possibility of performing experiments in quantum magnonics which might allow single localised magnon states to be created and measured. So far, the strong coupling regime of quantum electrodynamics has been reached\cite{Zhang2014,Huebl2013,Tabuchi2014,Goryachev2014a} along with demonstrations of magnetically induced transparency\cite{Zhang2014}. The strong coupling has been enabled by the high moment density and low magnetic damping\cite{Spencer1959} in YIG. Both uniform and non-uniform modes have shown strong coupling\cite{Goryachev2014a}. The work reported in this Letter has been performed in such a context. 

We fabricate an easily made cavity (Fig.~1a) with a well-defined non-uniform field specifically so that we can couple into the non-uniform excited modes. It is made from a short ($L=28$ mm) length of 3.5 mm diameter copper semi-rigid coaxial cable cut flat at each end. These ends are brought into proximity with similarly flat ends in connectorised leads, with a small air gap forming the coupling capacitance. SMA screw connectors provide mechanical stability and allow the size of the air gap, and hence the coupling capacitance, to be varied in a controlled way. At one extreme, the coaxial cables can be brought into contact with each other, transforming the cavity back into a transmission line. We find that the internal quality factor ($Q$) of our cavity is 515, in close agreement with the theoretical value of $Q=517$ calculated from the specified attenuation in the co-axial cable. For the cavity experiments described in this Letter, we tuned the coupling strengths to be $\kappa_c/2 \pi= 3.3$ MHz, giving a loaded $Q$ of 261, a fundamental frequency of $\omega_0/2\pi= 3.535$ GHz and a total cavity linewidth of $(2\kappa_c+\kappa_{int})/2 \pi =13.5$ MHz.

\begin{figure}
\includegraphics{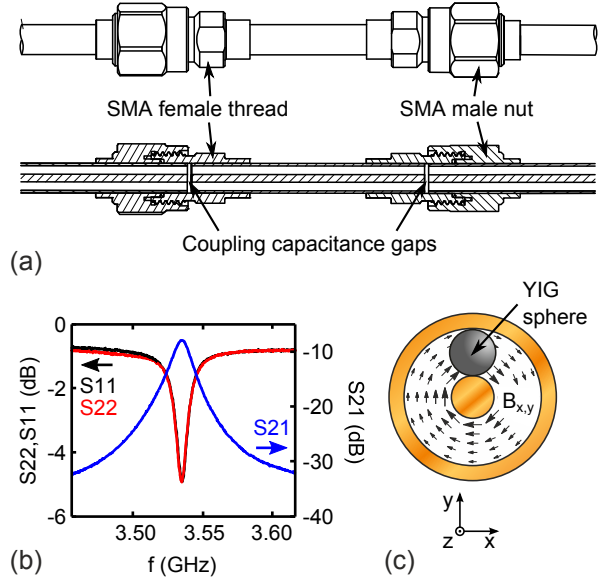}
\caption{\label{fig1}The cavity and YIG sphere. (a) Diagram and longitudinal cross-section of the cavity. It is made from 3.5 mm diameter (UT141) semirigid coaxial cable, and the gap capacitances controlled with SMA coupling threads. (b) $|\mathrm{S}21|$, $|\mathrm{S}11|$ and $|\mathrm{S}12|$ for the cavity configuration used in this experiment. (c) Non-uniform magnetic field around the YIG sphere due to the alternating cavity drive. The global field is applied in the $z$ direction.}
\end{figure}

A YIG sphere\footnote{Ferrisphere, Inc.} of diameter 1 mm is inserted into the cable dielectric at the midpoint of the cavity (Fig.~1c). A key feature of our cavity is the well defined and non-uniform magnetic field profile in the dielectric gap, which has a $1/r$ form in the radial direction. This non-uniform field allows the cavity to couple to both uniform and non-uniform spin-wave modes.

We measure the transmission, S21, of the system using a vector network analyser. The incident power on the cavity is -10 dBm; the driven FMR in this regime is linear, as observed by the independence of S21 on power. We sweep the frequency from 2 GHz to 8 GHz, encompassing both the fundamental mode and the second harmonic of the cavity. A magnetic field is applied parallel to the cavity, and is varied between 50 and 330 mT. In this field range the magnetization of the YIG is fully saturated.

\begin{figure}
\includegraphics{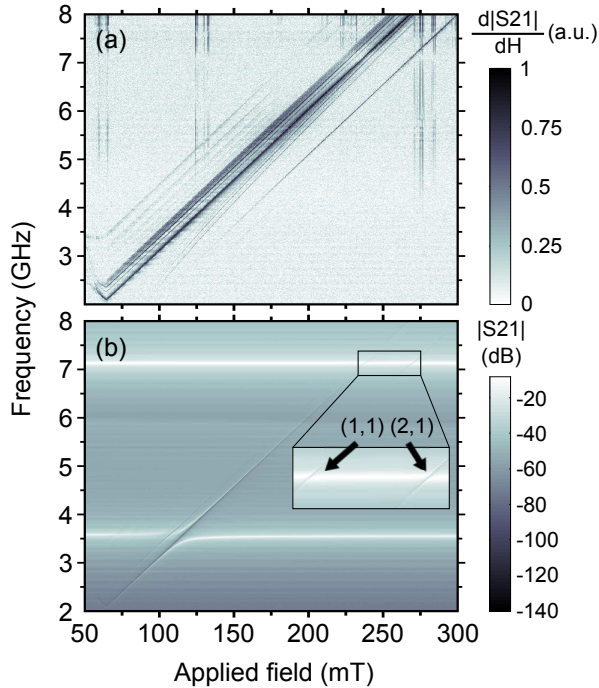}
\caption{\label{fig2}Transmission of the system. (a) Derivative of cavity transmission amplitude, $d|\mathrm{S}21|/dH$, with both coupling capacitors shorted; it acts as a $50$ $\Omega$ transmission line. Many magnetostatic modes are visible. (b) Transmission amplitude $|\mathrm{S}21|$ of the cavity with both coupling capacitances set to $\approx 28$ fF. Anticrossings between cavity modes and magnetostatic modes are seen. The coupling depends strongly upon which magnetostatic mode is being excited. The anticrossing between the (2,1) mode and the second cavity harmonic is labelled.}
\end{figure}

The transmission of the system is shown in Fig.~\ref{fig2}. In Fig.~\ref{fig2}a we show $d|\mathrm{S}21|/dH$ for the case in which the coupling capacitors are shorted; this is therefore simply transmission line FMR. The magnetostatic band can be clearly seen, comprising a multitude of modes. Unambiguous identification of each one is not trivial; the intensity of each line depends on both the coupling of the magnetostatic mode to the transmission line, and the damping of that mode\cite{Bilzer2007a}, and the linewidth is also dependent on the measurement method\cite{Kalarickal2006}.

In Fig \ref{fig2}b we revert to the gap coupled cavity as earlier described. Anticrossings between magnetostatic modes and the cavity resonances at both $3.53$ GHz and $7.12$ GHz are seen, with a maximum coupling strength of 130 MHz for the uniform FMR mode and the fundamental cavity frequency. Coupling to the second harmonic of the cavity is in general much weaker, as the sphere is positioned at a magnetic field node of this cavity mode.
	
The spatial form and resonant frequencies of modes in magnetized spheres is well known\cite{Walker1957,Fletcher1959}. Following Walker\cite{Walker1957} we label them with indices $n$ and $m$\footnote{For some modes, the resonance equation admits more than one solution, which is generally labelled with a third index. As none of the modes we explicitly discuss here have multiple resonances, for simplicity we omit this index.}. The radial form of the mode is characterized by $n$, and $m$ determines the number of lobes in the mode pattern.

The coupling of the $(n,m)$ mode to the cavity is given by\cite{Tabuchi2014}

\begin{align*}
g_j = \frac{\eta_{n,m}}{2} \gamma \sqrt{\frac{\hbar \omega_c \mu_0 \epsilon_r}{V_c}}\sqrt{2 N s}.
\end{align*}

Here $\omega_r$ is the resonance frequency, $V_c$ is the volume of the cavity mode, $N$ is the total number of spins in the YIG sphere, $s = 5/2$ is the spin per site, $\mu_0$ is the permeability of free space and $\epsilon_r$ is the relative permitivity of the dielectric within the co-axial cable. The overlap between the cavity mode and the sphere mode ($n,m$) is described by $\eta_{n,m}$, which is given by

	%

\begin{align*}
\eta_{n,m}=&\bigg\vert\frac{1}{H_{\textrm{max}} M_{\textrm{max}} V_s}\times \int_\textrm{sphere} (\mathbf{H}\cdot \mathbf{M})\; dV\bigg\vert.
\end{align*}
	
$\mathbf{H}$ is the r.f.~driving field, and $\mathbf{M}$ is the complex time-dependent off $z$ axis sphere magnetization for mode $(n,m)$. $H_{\textrm{max}}$ and $M_{\textrm{max}}$ are the maximum magnitudes of these, and $V_s$ is the sphere volume. The coupling strength is independent of magnetostatic damping.

The coupling to a particular FMR mode is dependent on the relative symmetries of the mode and the r.f. drive field. It is forced to zero if the mode is antisymmetric with respect to the drive. In particular, for the coupling to the fundamental cavity mode to be significant the FMR mode must be symmetric and low-order in $z$ (as the cavity mode is also symmetric). This condition is only met by modes for which $n=m$. In contrast, in order to couple to the second harmonic cavity mode, the mode must be antisymmetric about $z=0$. We tabulate calculated coupling constants larger than 1 MHz in Table \ref{table1}.

\begin{figure}
\includegraphics{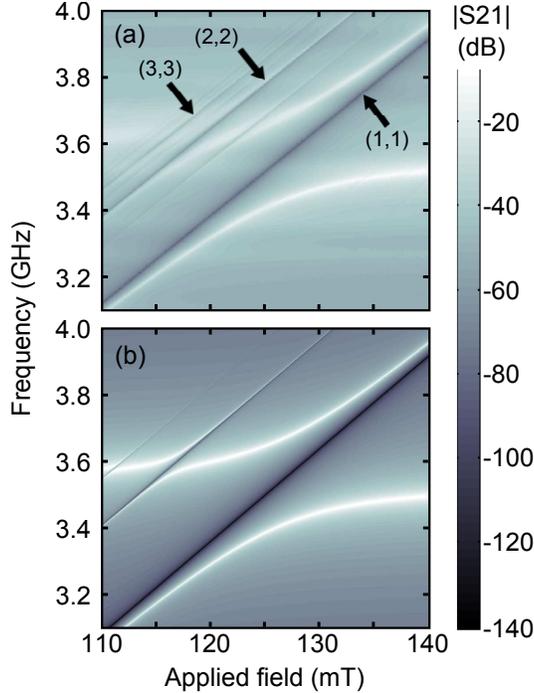}
\caption{\label{fig3}Strong coupling betwen cavity and FMR modes. (a) The region around the anticrossing of the uniform mode and the fundamental mode of the cavity. The most strongly coupled modes are labelled. (b) Simulation of the same region using the input-output formalism.}
\end{figure}

\begin{table}[h]
\caption{\label{table1}Calculated coupling strengths of selected FMR modes to the fundamental and second harmonic cavity resonances.}
\centering
\begin{tabular*}{\columnwidth}{@{\extracolsep{\fill}}cccc}
\hline \hline
\rule{0pt}{2.6ex}\rule[-1.2ex]{0pt}{0pt}&&\multicolumn{2}{c}{$g/2\pi$ (MHz)} \\ \cline{3-4}
\rule{0pt}{2.6ex}\rule[-1.2ex]{0pt}{0pt}~~~~$n$& $m$ & Fundamental & Second harmonic \\ \hline 
~~~~1 & 1 &130 & 0 \\ 
~~~~2 & 1 & 0 & 2.9 \\
~~~~2 & 2 & 27.1 & 0 \\
~~~~3 & 3 & 8.1 & 0 \\ 
~~~~4 & 4 & 2.8 & 0 \\ 
~~~~5 & 5 & 1.1 & 0 \\ \hline \hline

\end{tabular*}
\end{table}

In order to compare these values to our measurement we model the transmission of strongly coupled cavity using the input-output formalism\cite{Clerk2010,Zhang2014,Huebl2013,Tabuchi2014}. Close to the fundamental mode of the cavity

\begin{align*}
\text{S21}&= \nonumber \\
&\frac{\kappa_c}{i (\omega -\omega_c)-\frac{1}{2} (2\kappa_c+\kappa_{\textrm{int}})+\sum _j \frac{|g_j|^2}{-\frac{1}{2}\gamma_j+i (\omega -\omega _j)}},
\end{align*}

where $j$ runs over the magnetostatic modes and $\gamma_j$ are the FMR linewidths. In Fig.~3 we examine the region around the uniform mode's anticrossing with the cavity fundamental more closely.  In Fig.~3a we show the measured transmission, and in Fig.~3b show the calculated transmission over the same range. For $m=n$ modes the two are in good agreement. We attribute the appearance of additional weakly coupled modes to the YIG sphere being slightly off-center in the cavity, which lifts the symmetry conditions described above. This also accounts for the weak coupling of the uniform mode to the second harmonic of the cavity.

In conclusion, we have described a simple tunable cavity-spin ensemble system which can nevertheless achieve the strong coupling limit due to the high spin density in ferrimagnetic YIG. We show that the coupling to the uniform mode is 130 MHz, giving a cooperativity of $C=g^2/\kappa\gamma\approx 200$. Furthermore, the asymmetric but well defined field profile in the cavity permits a quantitative understanding of the coupling to higher order spin wave modes. Coupling between microwave cavities and highly tunable magnonic excitations is a candidate building block for hybrid quantum systems, and the ability to selectively excite specific spin wave modes offers intriguing possibilities in the emerging field of quantum magnonics.

We would like to acknowledge support from Hitachi Cambridge Laboratory, and EPSRC Grant No. EP/K027018/1. A.J.F. is supported by a Hitachi Research fellowship.

\bibliography{ResonatorCoupling}

\end{document}